\documentclass{article}
\usepackage{multicol}
\usepackage{cite}
\usepackage{amsmath,amssymb,amsfonts}
\usepackage{algorithmic}
\usepackage{graphicx}
\usepackage{textcomp}
\usepackage{xcolor}
\usepackage{algorithm2e}
\usepackage{pgfplots}

\usepackage{subfig}
\newcommand{\ignore}[1]{}
\ignore{
\def\BibTeX{{\rm B\kern-.05em{\sc i\kern-.025em b}\kern-.08em
    T\kern-.1667em\lower.7ex\hbox{E}\kern-.125emX}}
}

\title{Heuristic Random Designs for Exact Identification of Defectives Using Single Round Non-adaptive Group Testing and Compressed Sensing}

\author{Catherine Haddad-Zaknoon \\ Technion - Israel Institute of Technology \\ catherine@cs.technion.ac.il}
\date{}

\begin{document}

\maketitle
\begin{abstract}
Among the challenges that the COVID-19 pandemic outbreak revealed is the problem to reduce the number of tests required for identifying the virus carriers in order to contain the viral spread while preserving the tests reliability. To cope with this issue, a prevalence testing paradigm based on group testing and compressive sensing approach or \emph{GTCS}  was examined. In these settings, a non-adaptive group testing algorithm is designed to rule out sure-negative samples. Then, on the reduced problem, a compressive sensing algorithm is applied to decode the positives without requiring any further testing besides the initial test matrix designed for the group testing phase. The result is a single-round non-adaptive group testing - compressive sensing algorithm to identify the positive samples.
  
In this paper, we propose a heuristic random method to construct the test design called \emph{$\alpha-$random row  design} or $\alpha-$RRD. In the  $\alpha-$RRD, a random test matrix is constructed such that each  test aggregates at most $\alpha$ samples in one group test or pool. The pooled tests are heuristically selected one by one such that samples that were previously selected in the same test are less likely to be aggregated together in a new test. We examined the performance of the $\alpha-$RRD design within the GTCS paradigm for several values of $\alpha$. The experiments were conducted on synthetic data. Our results show that, for some values of $\alpha$, a reduction of up to 10 fold in the tests number can be achieved when $\alpha-$RRD design is applied in the GTCS paradigm.  

\textbf{keywords} Group Testing, Pooling Design, Compressive Sensing, COVID19-PCR

\end{abstract}

\section{Introduction}
The problem of \emph{group testing} is the problem of identifying a small amount of  \emph{items} or \emph{subjects} known as  \emph{defective items} or \emph{positive subjects} within a pile of elements using \emph{group tests} or \emph{pools}.  Let $X=[n]:=\{1,\cdots , n\}$ be a set of $n$ items or \emph{subjects} to be tested, and let $I\subseteq X$ be the set of positive subjects such that  $|I| = d < n$. A \emph{group test}  or a \emph{pool} is a subset $Q\subseteq X$ of subjects. A test result is \emph{positive} if $Q\cap I \neq \emptyset$ and \emph{negative} otherwise. Alternatively, we identify the test $Q\subseteq X$ with an {\it assignment} $a\in\{0,1\}^n$ where $a_i=1$ if and only if $i\in Q$, i.e., $a_i=1$ signifies that item $i$ participates in the test $Q$.  The objective of group testing algorithms is to find the members of the set $I$ with minimum number of group tests.

Group testing algorithms  can be deterministic or randomized, \emph{adaptive}  or \emph{non-adaptive}. In  adaptive algorithms, the tests might be selected based on the results of previous ones. In non-adaptive algorithms, tests are independent and must not rely on previous results. Therefore, all the tests can be performed in a single parallel step. The set of tests in any non-adaptive deterministic (resp. randomized) algorithm can be identified with an (resp. random) $m\times n$ {\it test design matrix} $M$ (also called \emph{pool design}) that its rows are all the assignments $a$ that correspond to the group tests sets selected by the algorithm. The group testing algorithm considered in this work is random non-adaptive.

Group testing approach was first introduced during  WWII~\cite{Dor43}, when Robert Dorfman, in 1943,  suggested the method to reduce the expected number of  tests needed to weed out all syphilitic soldiers in a specific unit. Since it was first suggested, group testing algorithms have been diversely applied in a wide range of areas. These applications include DNA library screening~\cite{DH2_00, DH06}, product quality control testing \cite{SG59}, file searching in storage systems~\cite{KS64}, sequential screening of experimental variables~\cite{Li62}, efficient contention algorithms for MAC~\cite{KS64,JW85}, random access communications problem \cite{JW85}, data compression~\cite{HL02}, and computations in data stream model~\cite{CM03}. For a brief history and other applications, the reader is referred to~\cite{Ang88,Ci13,DH00,DH06,H72,MP04,DH2_00,KS64,HL02,CM03} and the references therein.

Among its recent applications, due to the recent pandemic outbreak, group testing approach for accelerating COVID-19 testing was widely applied across the globe.  Due to severe shortages in testing kits supply, a number of researches adopted the group testing paradigm for COVID-19 mass testing not only to accelerate the testing process, but also to  reduce the number of the tests required to reveal positive virus-carriers~\cite{YK20,GC20, MCR20, SHGOY20, CSPL20, EHWB20, BENAMI2020}.   In many labs, COVID-19  detection was performed using PCR methods. PCR-based machines can perform multiple parallel tests in single run, while each run can be several hours long. Driven by the process of PCR testing, non-adaptive group testing is most fit for these settings. In this context, the items in question are samples taken from potential patients and the \emph{positive subjects} are samples that test positive to the virus. 

While many researchers applied Dorfman's attitude with multi-stage PCR runs, some have examined the possibility of using the method while designing tests that require no more than one PCR-round. One of the promising directions is the group testing - compressed sensing paradigm  (GTCS) used in~\cite{Ghosh2020, Ghosh2020CS, YMX, SNSWVS20}. This method includes the following stages; initially, a test  matrix $M$ is designed for a single non-adaptive group testing round. Upon test results delivery, a two-stage decoding process is performed. The decoding process is purely combinatorial and does not involve any further sample testing. Using standard non-adaptive group testing decoding (e.g. COMP algorithm~\cite{COMP}), a substantial amount of samples that tested negative to the virus are ruled out. Obviously, the main benefit of this phase is to reduce the dimension of the compressed sensing problem by cutting down the number of samples that need further decoding. This is crucial due to the computational complexity of compressed sensing algorithms. In the next stage, compressive sensing techniques are used over the reduced problem (e.g. Orthogonal Matching Pursuit - OMP~\cite{OMP} , Fast-OMP~\cite{FOMP}), to analyze and identify real carriers.

The design of the test matrix $M$ is crucial for both group testing phase and the compressive sensing phase that follows. In ~\cite{SNSWVS20}, the  design matrix is constructed using Reed Solomon error correcting codes. The authors has checked their method on a set of $n=384$ samples in which $5$ samples are positive (about $1.3\%$). For pool size of $48$ and using $48$ group tests they could recover all the $5$ positive samples. In the work of Jirong, Mudumbai and Xu. ~\cite{YMX}, the authors investigated two types of pooling designs. The first is Bernoulli random matrix where each entry is selected to be $1$ or $0$ with equal probability. The second design is obtained using expander graphs where each column has a fixed number of non-zero entries. The designs tested in \cite{Ghosh2020CS} are based on Kirkman triples.  

In this paper, we propose a heuristic random method to construct the test design $M$ called \emph{$\alpha-$random row  design} or $\alpha-$RRD. In the  $\alpha-$RRD, a random test matrix is constructed such that each  test in $M$ aggregates at most $\alpha<n$ samples in one group test. This model is useful in applications were tests reliability might be compromised if the pool size is large.  The matrix rows are selected one by one.  The main idea of the construction is to choose the non-zero entries of a new row according to two considerations:  samples that belong to the same subject participate in similar number of tests on average (fairness); and samples that were previously selected in the same test are less likely to be aggregated together in a new test (sparsity). We perform experiments on noiseless synthetic data to examine the performance of the design, while applying OMP (Orthogonal Matching Pursuit) as the compressive sensing algorithm. Practically, test designs need to be deterministic, meaning, they need to be predefined before the testing process. To use random test design, it is acceptable to make simulations of several random designs and choosing the design that performs best on some set of data. Then, this design is adopted to be used as a deterministic one for the real time tests.

Our experiments results suggest that using the GTCS framework with $\alpha-$RRD design can reduce the number of tests dramatically. First, the results imply that there is an evident correlation between the value of $\alpha$ and the performance of the process; choosing higher values of $\alpha$ can increase the success rate in identifying the positives. This can be observed from the results summarized in Fig.~\ref{m96}. For example, for $n=400$ and positives rate $5\%$, the success rate for $\alpha=10$ is bellow $20\%$ while it can be increased beyond $95\%$ when $\alpha=22$. This paradigm is reproduced in all test configurations  outlined in Figure(\ref{m96}). In the context of COVID-19 testing, the term \emph{positives rate} or \emph{positives ratio} indicates the rate of positives within some community.  The simulations show that the number of tests can be reduced by up to $10$ folds, relative to testing samples one by one, when  $\alpha =48$ for $n=900$ and positives rate of $1\%$. Table~(\ref{m96}) summarizes the performance of the $\alpha-$RRD design when the number of the tests is fixed to $m=96$ for positives rates of $1\%,\ldots ,5\%$ and $n=400, \ldots, 900$. An additional set of experiments examines the performance of the design when $m=384$, positives rates of $1\%, \ldots, 10\%$ and the number of samples $n$ that reaches $4000$. Similar results are obtained in these dimension settings as summarized in Table~(\ref{m384bestrate}). The values of $m$ are derived from practical considerations that involve the common sizes of PCR-based machines commonly used in the industry.

\section{Definitions and Preliminaries}

\subsection{The group testing  $-$ GT problem }
Let $X=[n]:=\{1,\cdots , n\}$ be a set of $n$ items or subjects, and let $I\subseteq X$ be the set of positive (defective) items such that  $|I| = d\ll n$. A \emph{group test}  or a \emph{pool} is a subset $Q\subseteq X$ of items. The quantity $p:=|Q|$ is called the \emph{pool size}. The result of the test $Q$ with respect to $I$ is defined by $Q(I):=1$ if $Q\cap I \neq \emptyset$ and $Q(I):=0$ otherwise. Alternatively, we identify the test $Q\subseteq X$ with an {\it assignment} $a\in\{0,1\}^n$ where $a_i=1$ if and only if $i\in Q$. 

The set of tests in any non-adaptive group testing algorithm can be identified with an $m\times n$ {\it test design matrix} $M$ (pool design) where each row corresponds to  an assignment $a\in \{0,1\}^n$ that defines a group test selected by the algorithm. Upon performing the tests defined by $M$, each test of the $m$ assignments in $M$ yields the value $1$ or $0$ according to whether the tests contains at least one positive sample or not.  Let $y\in \{0,1\}^m$ denote the test results obtained by performing the tests of $M$, and let $x\in \{0,1\}^n$ be a vector such that $x_i=1$ if and only if $i\in I$.  Formally,
$$y=M\odot x,$$
where the operation $\odot$ is defined as follows; for each $1 \leq i \leq m$,
\begin{equation}
y_i = \bigvee_{j=1}^n M_{i,j}\cdot x_j,
\label{YIGT}
\end{equation}
where the $\vee$ operation is the logic OR. It is easy to see that the definition from ~\ref{YIGT} is equivalent to $y_i=1$ if and only if $M_{(i)}\cap I\neq \emptyset$, where $M_{(i)}$ is the set that corresponds to the test defined by the $i$th row in $M$.

\subsection{The compressive sensing $-$ CS problem }
Assume that each subject sample can be measured by a real valued number that expresses the \emph{magnitude} or the \emph{load} of the examined symptom (e.g. viral load in COVID-19 case).  Let $\hat{x}\in R^n$ be a $n-$dimensional real-valued vector that signifies the symptom load of the subjects; i.e. for each $1 \leq i \leq n$, $\hat{x}_i$ indicates the symptom load of the subject $i$, where the value of $\hat{x}_i$ is directly proportional to the load. Symptom-free items will have their corresponding load measure equals to $0$.  We assume that the number of positives $d \ll n$, therefore, the load vector $\hat{x}$ is $d-$sparse, meaning, it includes only $d$ non-zero entries. The objective is to restore the indexes of the non-zero entries in $\hat{x}$. 

Similar to the definition of the result vector $y$ from the GT settings, the design matrix $M$, also called the \emph{sensing matrix} in the compressive sensing context, defines the load vector $\hat{y} \in R^n$ where each entry $\hat{y}_i$  correlates with the load of the $i$th pool in $M$. That is,
\begin{equation}
\hat{y} = M\cdot \hat{x},
\label{YCS}
\end{equation}
where the $(\cdot)$ operation is the standard matrix multiplication, therefore, for each $1 \leq i \leq m$, 
 \begin{equation}
 \hat{y}_i = \sum_{j=1}^{n}M_{i,j}\cdot \hat{x}_j. 
 \label{YICS}
 \end{equation}

In this paper, we will be interested in restoring the indexes of the non-zero entries of the vector $\hat{x}$, which is equivalent to restoring the binary vector $x$ from the GT settings.

\subsection{The group testing - compressive sensing paradigm - GTCS}
The GTCS paradigm suggests a non-adaptive group testing generic algorithm for identifying the exact set of positives while using CS-based decoding techniques. The GTCS paradigm is composed of three basic phases. 

\begin{enumerate}
\item \textbf{Create and perform the actual tests:} Create a test design $M$ and perform the group tests defined by the design. Practically, the outcome of this stage is a vector $\hat{y}\in R^n$ as described in Equations (\ref{YCS}) and (\ref{YICS}). This stage is followed by two-stage decoding process to exactly identify the test of positives. The vector $y$ is derived from $\hat{y}$ by assigning  each entry $y_i=1$ if $\hat{y}_i >0$ and $y_i=0$ otherwise.  
\item \textbf{Group testing decoding:} using standard group testing decoding methods (e.g. COMP algorithm~\cite{COMP}) on the problem $y=M\odot x$,  a subset $X_0\subseteq X$ of items that are guaranteed by the GT algorithm to be negative samples is identified. This stage is used to reduce the size of the problem to be solved in the next stage. The rational behind this step is to exploit the fact that GT decoding algorithms like COMP has zero false negatives (i.e. all  sample that were detected by the algorithm as negative ones are actually negative). Therefore, eliminating the set of sure-negative samples $X_0$ reduces the computational complexity of the step that follows, while keeping its decoding accuracy intact. 

The reduced compressive-sensing problem is established by applying the following enhancements. Given the set $X_0$ that include sure negatives, we define a new set $X_r:= X \setminus X_0$. Let $Y_0\subseteq [m]$ be the set of tests indexes that yielded the result $0$ in the previous stage. The new test design matrix $M_r$ is constructed from $M$, $X_0$ and $Y_0$ by projecting $M$ on the columns that correspond to the samples in $X\setminus X_0$ and the rows that appear in the set $[m]\setminus Y_0$. Therefore, the resulting matrix is an $(m_r\times n_r)$ binary matrix where $m_r = m-|Y_0|$ and $n_r=n-|X_0|$. The reduced test result vector $\hat{y}_r$ is derived from $\hat{y}$ by deleting the entries that correspond to $Y_0$. 
\item \textbf{Compressive sensing decoding:} by applying standard compressive sensing algorithms (e.g. OMP, LASSO) on the reduced problem $\hat{y}_r = M_r\cdot \hat{x}_r$, and using the results from previous stage, the vector $\hat{x}_r$ and therefore, the vectors $\hat{x}$ and $x$ can be restored.   
\end{enumerate}

\section{Random Row Design - $\alpha-RRD$}

In this section, we propose a random design for GT. For this design, we restrict that the pool size to be at most $\alpha<n$. The design is constructed row by row. The main idea of the construction is to choose the non-zero entries in the new row according to two principles;
\begin{enumerate}
\item \textbf{Fairness:} Elements that participated in the minimum number of tests in previous rows will be more likely to be chosen in the new test.
\item \textbf{Sparsity}: Elements that were previously selected in the same test will be less likely to be assembled together in the new test.
\end{enumerate}

For a vector $a=(a_1, \ldots, a_n)\in \{0,1\}^n$, recall that ${\cal{H}}(a):=\{i: a_i =1\}\subseteq[n]$. The \emph{Hamming weight} of $a$ is denoted by $\omega(a)$ and is equal to $\omega(a) = |{\cal{H}}(a)|$. Let $\mathbf{0}^n$ ($\mathbf{1}^n$, $\mathbf{\infty}^n$) denote the all zero (one, $\infty$ resp.) vector of length $n$. Let $A_{(m\times n)}$ be an $m\times n$ matrix over $\{0,1\}$. For all $1\leq i\leq m$,  denote by $A_{(i)}\in \{0,1\}^n$ the $i$th row of the matrix $A$, by $A^{(j)}$ the $j$th column and by $A_{i,j}$ the element in $A$ that corresponds to the $i$th row and the $j$th column. The \emph{columns weight vector} of a matrix $A_{m\times n}$ is a vector $w=(w_1,\ldots, w_n)\in R^n$ such that $w_j = \sum_{i=1}^{m}A_{i,j}$.  Practically, the weight column vector indicates the number of tests each element participated in.

The procedure $\mathbf{RRD}(n,m,\alpha)$ in Fig.~\ref{RRSDAlg} describes the RRD strategy to choose a random design $A$ with $m$ rows and $n$ columns where each row is of Hamming weight at most $\alpha$. The algorithm starts by randomly choosing the first row in $A$  from the set of binary vectors of length $n$ and Hamming weight  $\alpha$. Assume that the first $\ell-1$ rows are already chosen, and let $A_{\ell -1}$ be the matrix defined by those rows. Let $\hat{w} = (w_1, \ldots, w_n)\in R^n$ be the columns weight vector of $A_{\ell-1}$. Then, the algorithm chooses the first non-zero entry in the $\ell$th row uniformly randomly from the set of indexes that correspond to the entries of minimal value in $\hat{w}$.  This choice complies with the fairness principle.

Let $k <\alpha$ be the number of non-zero entries that algorithm already chose for the $\ell$th row. The $k+1$ entry is chosen as follows. Let $Q_k$ be the set of indexes of the non-zero entires chosen so far in the current row. Let $Z$ be the set of rows indexes $i$,  such that ${\cal{H}}(A_{(i)})\cap Q_k \neq \emptyset$, and let $\hat{w}$ be the weight vector of submatrix of $A$ defined by the rows in $Z$. The algorithm evaluates $\hat{w}$ in steps 2.1.2 and 2.1.3 in Fig.~\ref{RRSDAlg}.  The algorithm then constructs the set of indexes $S\subseteq [n]$ that includes all the indexes $j$ such that $w_j$ is of minimum value among the entries in $\hat{w}$ and sums the corresponding columns. Let $X$ be the set of column indexes with minimum value. Then, among the indexes in X, choose $s\in X$ uniformly at random and assign $A_{\ell, s} = 1$. These are steps 2.1.4 $-$ 2.1.11 in Fig.~\ref{RRSDAlg}. This choice complies with the fairness principle.

The procedure $\mathbf{CalcSelectedRows}$ from Fig.~\ref{PROCD}.(c) outputs a set of row numbers $C\subseteq \{1, \ldots, \ell-1\}$ such that $i\in C$ if and only if ${\cal{H}}(A_{(i)})\cap Q_k\neq \emptyset$. The procedure $\mathbf{UpdateWeight}$ calculates the weight vector $w =\sum_{j\in C}A_{(j)}$ (See step no. 2 in Fig.~\ref{PROCD}.(a)). To ensure that the entires in $Q_k$ are excluded from the selection of the next non-zero index, the weight vector $w$ is updated to have the value $\infty$ in the corresponding indexes. (See step 3 in Fig.~\ref{PROCD}.(a)).

The selection of the set $S$ complies with the sparsity principle. The set $S$ is derived from the weight vector $\hat{w}$ by selecting the indexes with minimal weight, where the weight is evaluated over the rows that agree with one or more of the entries selected for the current test. The initialization step in $\mathbf{SumColumns}$ implies that the choice of the next non-zero entry of the current test will be from the indexes in $S$.  Those indexes are the ones with minimum agreement with the current test, therefore, choosing the next non-zero entry from them is the best choice to keep the sparsity principle. The selection of the set $X$ in step 2.1.8 complies with the fairness principle; among the best candidates from the indexes of $S$, the algorithm chooses $s$ uniformly from those that appeared minimum number of times over all the previous tests.

\color{black}
\begin{figure*}[t]
  \begin{center}
  \fbox{\fbox{\begin{minipage}{18em}
  \begin{tabbing}
  xxxx\=xxxx\=xxxx\=xxxx\= \kill
    {{$ \mathbf{RRD}$} $(n,m, \alpha)$}\\
  
 1) $A \leftarrow \{0\}_{(m\times n)}.$\\
  2) Choose $a\in \{0,1\}^n$ uniformly at random from all binary vectors of weight $\alpha$. \\
  3) $A_{(1)}\leftarrow a$. \\
  4)  For each $\ell=2,\ldots, m$ do:\\
  \> 2.1)  $k\leftarrow 0$, $Q_0\leftarrow \emptyset$\\
  \>2.1) while $ (k < \alpha) $\\
      \> \>2.1.1) $k\leftarrow k+1$ \\ 
   \>\>2.1.2) $C\leftarrow \mathbf{CalcSelectedRows}(n,Q_{k-1}, A,\ell-1) $. \\
    \> \> 2.1.3) $\hat{w}\leftarrow \mathbf{UpdateWeight}(n,Q_{k-1},  C, A)$\\
  \> \>2.1.4) ${\hat{w}}_{\min}\leftarrow \min_{1 \leq j\leq n}\hat{w}_j$\\
  \> \>2.1.5) $S\leftarrow \{p: \hat{w}_p = \hat{w}_{\min} \}$\\
   \>\>2.1.6) $\hat{z}\leftarrow \mathbf{SumColumns}(n,A, \ell-1, S)$ \\ 
   \> \>2.1.7) ${\hat{z}}_{\min}\leftarrow \min_{1 \leq j\leq n}\hat{z}_j$\\
  \> \>2.1.8) $X\leftarrow \{t: \hat{z}_t = \hat{z}_{\min} \}$\\
  \> \>2.1.9) Select $s$ uniformly at random from $X$.\\
  \> \>2.1.10) $Q_k\leftarrow Q_{k-1}\cup \{s\}$.\\
  \>\>2.1.11) $A_{\ell, s} \leftarrow 1$\\
   
5) Output ($A$).
    \end{tabbing}
  \end{minipage}}}
  \end{center}
	\caption{Procedure $\mathbf{RRD}$ for creating a random design $A$} 
	\label{RRSDAlg}
	\end{figure*}

\ignore{
\begin{figure}[t]
\centerline{\includegraphics{fig1.png}}
\caption{Example of a figure caption.}
\label{fig}
\end{figure}
}
	
\begin{figure*}[htp] 
    \centering
       \subfloat[$\mathbf{UpdateWeight}$ procedure]{%
      \fbox{\fbox{\begin{minipage}{28em}
  \begin{tabbing}
  xxxx\=xxxx\=xxxx\=xxxx\= \kill
    {{$ \mathbf{UpdateWeight}$} $(n,Q, C, A)$}\\

  1) If $(C=\emptyset)$\\
  \>  $w\leftarrow \mathbf{1}^n$. \\
  2) Else\\
  \>  $w\leftarrow \sum_{j\in C}  A_{(j)}$ \\ 
   3) For  $i=1,\ldots, n$ do:\\
  \> 3.1) If $(i \in Q)$, then \\
  \>\> $w_i\leftarrow \infty$ \\
  4) Output ($w$) \\  
    \end{tabbing}
  \end{minipage}}}
         \label{fig:a}%
        }%
    \hfill%
    \subfloat[$ \mathbf{SumColumns}$ procedure]{%
 \fbox{\fbox{\begin{minipage}{28em}
  \begin{tabbing}
  xxxx\=xxxx\=xxxx\=xxxx\= \kill
    {{$ \mathbf{SumColumns}$} $(n,A,\ell,  S)$}\\
  1) $w\leftarrow \mathbf{\infty}^n$ \\
  2)  For each $i\in S$ do:\\
  \> 2.1)  $w_i\leftarrow \sum_{j=1}^{\ell}A_{j,i}$\\
  3) Output ($w$) \\
    \end{tabbing}
  \end{minipage}}}
        \label{fig:a}%
        }%
    \hfill%
    \subfloat[$ \mathbf{CalcSelectedRows}$ procedure]{%
  \fbox{\fbox{\begin{minipage}{28em}
  \begin{tabbing}
  xxxx\=xxxx\=xxxx\=xxxx\= \kill
    {{$ \mathbf{CalcSelectedRows}$} $(n,Q, A, \ell)$}\\
  1) $C\leftarrow \emptyset$ \\
  2)  For each $i=1,\ldots,\ell$ do:\\
  \> 2.1) If $({\cal{H}}(A_{(i)})\cap Q \neq \emptyset)$, then\\
  \>\>  $C\leftarrow C\cup \{i\}$ \\
  3) Output ($C$) \\
    \end{tabbing}
  \end{minipage}}}
        \label{fig:b}%
        }%
    \caption{Subroutines called from the $\mathbf{RRD}$ main procedure.}
    \label{PROCD}
\end{figure*}	
	
\label{RSSDSec}

\section{Experiments and Simulations}

In this section, we outline tests results of the performance of the $\alpha-$RRD design when chosen as the test design in the GTCS generic paradigm. \ignore{Although the construction of the  $\alpha-$RRD does not have any constraints on the dimensions of the design matrix, we have examined the new design on a predefined set of matrix dimensions dictated by  practical applications that involve PCR testing (e.g. rt-PCR testing for COVID-19).}  For the GT decoding, the COMP algorithm is selected to generate initial sure-negative set, while the OMP is used as the CS algorithm in the final stage. We performed the test on noiseless synthetic data sets.
\ignore{
In the second set of tests, we compare the performance of  $\alpha-$RRSD designs with other well-known random designs like RID, RrSD, RSsD and transversal design. The properties and set of parameters used in the construction of these designs are explained in the following subsections.
}

In this set of simulations, we examine the performance of the $\alpha-$RRD design for two major categories of settings. The first category fixes the number of the tests $m$ to be $96$ while the choice of $m$ in the second category is $m=384$. Despite that fact that the construction from section~\ref{RSSDSec} does not impose any limitation on the parameter $m$,  the choices of the values of $m$ in the experiments are derived from the number of tests that can be performed in parallel in most PCR machines used in the industry\ignore{~\cite{CITESOMETHINGABOUTPCRMACHINES}}. 

\begin{figure*}[htp] 
      
\fbox{\begin{tikzpicture}[scale=0.33]
\begin{axis}
	[
    title={(a). Identification rate for $n=400, m=96$},
    xlabel={\# Positives, $d$ },
    ylabel={Success rate\%},
    xmin=0, xmax=20,
    ymin=0, ymax=110,
    xtick={0,2,4,6,8,10, 12, 14, 16, 18, 20},
    ytick={0,10,20,30, 40,50, 60,70, 80,90,100},
    legend pos=south west,
    ymajorgrids=true,
    grid style=dashed,
    ]
\addplot    coordinates {
    (1,100)(2,100)(3,100)(4,100)(5,99)(6,98)(7,90)(8,91)(9,79)(10,76)(11, 68)(12, 63)(13,52)(14,52)(15,50)(16,25)(17,25)(18,21)(19,13)(20,17) };

  \addplot  coordinates {
    (1,100)(2,100)(3,100)(4,100)(5,99)(6,100)(7,97)(8,98)(9,96)(10,91)(11, 94)(12, 85)(13,81)(14,79)(15,73)(16,68)(17,65)(18,56)(19,43)(20,29) };
    
     \addplot   coordinates {
    (1,100)(2,100)(3,100)(4,100)(5,100)(6,100)(7,100)(8,100)(9,100)(10,98)(11, 96)(12, 98)(13,96)(14,96)(15,92)(16,87)(17,91)(18,81)(19,72)(20,64) };
  
  \addplot   coordinates {
    (1,100)(2,100)(3,100)(4,100)(5,100)(6,100)(7,100)(8,100)(9,100)(10,100)(11, 100)(12, 100)(13,100)(14,99)(15,98)(16,97)(17,92)(18,94)(19,86)(20,82) };
    
    \addplot   coordinates {
    (1,100)(2,100)(3,100)(4,100)(5,100)(6,100)(7,100)(8,100)(9,100)(10,100)(11, 100)(12, 100)(13,100)(14,99)(15,98)(16,98)(17,97)(18,98)(19,93)(20,89) };
    
 \addplot [color = orange, mark=triangle]  coordinates {
    (1,100)(2,100)(3,100)(4,100)(5,100)(6,100)(7,100)(8,100)(9,100)(10,100)(11, 100)(12, 100)(13,100)(14,99)(15,98)(16,99)(17,100)(18,100)(19,97)(20,94) };

  \addplot[color=green, mark=*]  coordinates {
    (1,100)(2,100)(3,100)(4,100)(5,100)(6,100)(7,100)(8,100)(9,100)(10,100)(11, 100)(12, 100)(13,100)(14,100)(15,100)(16,100)(17,100)(18,100)(19,99)(20,97) };

   \legend{$ \alpha=10$ , $\alpha=12$ , $\alpha=14$, $\alpha=16$, $\alpha = 18$, $\alpha=20$, $\alpha=22$}
   
\end{axis}
\end{tikzpicture}
\qquad}
\fbox{\begin{tikzpicture}[scale=.33]
\begin{axis}
	[
    title={(b). Identification rate for $n=500, m=96$},
    xlabel={\#Positives, $d$ },
    ylabel={Success rate\%},
    xmin=0, xmax=26,
    ymin=0, ymax=110,
    xtick={0,2,4,6,8,10, 12, 14, 16, 18, 20, 22, 24, 26},
    ytick={0,10,20,30, 40,50, 60,70, 80,90,100},
    legend pos=south west,
    ymajorgrids=true,
    grid style=dashed,
    ]
         
  \addplot  coordinates {
    (1,100)(2,100)(3,100)(4,97)(5,99)(6,92)(7,87)(8,84)(9,71)(10,67)(11, 63)(12, 47)(13,35)(14,24)(15,23)(16,14)(17,11)(18,10)(19,3)(20,4) (21,3)(22,1)(23,3)(24,0)(25,0)  };

 \addplot  coordinates {
    (1,100)(2,100)(3,100)(4,100)(5,100)(6,100)(7,99)(8,97)(9,98)(10,96)(11, 95)(12, 91)(13,90)(14,84)(15,74)(16,64)(17,66)(18,52)(19,48)(20,45) (21,33)(22,24)(23,14)(24,15)(25,9)  };

\addplot  coordinates {
    (1,100)(2,100)(3,100)(4,100)(5,100)(6,100)(7,100)(8,100)(9,100)(10,100)(11, 100)(12, 99)(13,98)(14,96)(15,94)(16,89)(17,85)(18,87)(19,71)(20,60) (21,65)(22,48)(23,46)(24,41)(25,25)  };

\addplot  coordinates {
    (1,100)(2,100)(3,100)(4,100)(5,100)(6,100)(7,100)(8,100)(9,100)(10,100)(11, 100)(12, 100)(13,100)(14,97)(15,99)(16,98)(17,96)(18,91)(19,84)(20,79) (21,77)(22,66)(23,56)(24,53)(25,38)  };
 \addplot [color = green, mark=*]  coordinates {
    (1,100)(2,100)(3,100)(4,100)(5,100)(6,100)(7,100)(8,100)(9,100)(10,100)(11, 100)(12, 100)(13,100)(14,100)(15,99)(16,99)(17,100)(18,94)(19,92)(20,80) (21,83)(22,79)(23,65)(24,59)(25,41)  };

 \addplot [color = orange, mark=*]  coordinates {
    (1,100)(2,100)(3,100)(4,100)(5,100)(6,100)(7,100)(8,100)(9,100)(10,100)(11, 100)(12, 100)(13,100)(14,100)(15,99)(16,100)(17,99)(18,99)(19,93)(20,91) (21,89)(22,79)(23,78)(24,66)(25,61)  };

 \addplot [color = cyan, mark=triangle]  coordinates {
    (1,100)(2,100)(3,100)(4,100)(5,100)(6,100)(7,100)(8,100)(9,100)(10,100)(11, 100)(12, 100)(13,100)(14,100)(15,100)(16,99)(17,97)(18,99)(19,96)(20,93) (21,88)(22,79)(23,73)(24,71)(25,63)  };

   \legend{ $\alpha=12$, $\alpha=16$, $\alpha=20$, $\alpha=24$, $\alpha=28$, $\alpha =32$, $\alpha=36$}
   
\end{axis}
\end{tikzpicture}
\qquad}
\fbox{\begin{tikzpicture}[scale=0.33]
\begin{axis}
	[
    title={(c). Identification rate for $n=600, m=96$},
    xlabel={\# Positives, $d$ },
    ylabel={Success rate\%},
    xmin=0, xmax=26,
    ymin=0, ymax=110,
    xtick={0,2,4,6,8,10, 12, 14, 16, 18, 20, 22, 24, 26},
    ytick={0,10,20,30, 40,50, 60,70, 80,90,100},
    legend pos=south west,
    ymajorgrids=true,
    grid style=dashed,
    ]
  \addplot  coordinates {
    (1,100)(2,100)(3,100)(4,96)(5,93)(6,93)(7,83)(8,74)(9,61)(10,53)(11, 46)(12, 37)(13,17)(14,24)(15,19)(16,13)(17,5)(18,3)(19,1)(20,0) (21,0)(22,0)(23,0)(24,0)(25,0)  };

 \addplot  coordinates {
    (1,100)(2,100)(3,100)(4,98)(5,100)(6,99)(7,93)(8,95)(9,91)(10,85)(11, 78)(12, 74)(13,59)(14,63)(15,48)(16,35)(17,35)(18,16)(19,24)(20,12) (21,8)(22,6)(23,4)(24,4)(25,1)  };

\addplot  coordinates {
    (1,100)(2,100)(3,100)(4,100)(5,100)(6,100)(7,100)(8,100)(9,97)(10,96)(11, 98)(12, 93)(13,88)(14,92)(15,82)(16,70)(17,69)(18,53)(19,46)(20,43) (21,33)(22,21)(23,17)(24,14)(25,4)  };

\addplot  coordinates {
    (1,100)(2,100)(3,100)(4,100)(5,100)(6,100)(7,100)(8,100)(9,100)(10,100)(11, 99)(12, 99)(13,94)(14,96)(15,94)(16,90)(17,81)(18,82)(19,63)(20,61) (21,47)(22,43)(23,36)(24,28)(25,14)  };
    
 \addplot [color = green, mark=*]  coordinates {
    (1,100)(2,100)(3,100)(4,100)(5,100)(6,100)(7,100)(8,100)(9,100)(10,99)(11, 100)(12, 99)(13,100)(14,97)(15,94)(16,95)(17,86)(18,85)(19,74)(20,67) (21,63)(22,51)(23,34)(24,25)(25,30)  };

 \addplot [color = orange, mark=*]  coordinates {
    (1,100)(2,100)(3,100)(4,100)(5,100)(6,100)(7,100)(8,100)(9,100)(10,99)(11, 100)(12, 100)(13,100)(14,99)(15,98)(16,93)(17,91)(18,88)(19,79)(20,72) (21,63)(22,46)(23,39)(24,33)(25,26)  };

 \addplot [color = cyan, mark=triangle]  coordinates {
    (1,100)(2,100)(3,100)(4,100)(5,100)(6,100)(7,100)(8,100)(9,100)(10,99)(11, 100)(12, 100)(13,100)(14,100)(15,97)(16,97)(17,95)(18,92)(19,85)(20,75) (21,66)(22,63)(23,61)(24,44)(25,29)  };

 \addplot   coordinates {
    (1,100)(2,100)(3,100)(4,100)(5,100)(6,100)(7,100)(8,100)(9,100)(10,99)(11, 100)(12, 100)(13,100)(14,100)(15,98)(16,94)(17,94)(18,87)(19,85)(20,77) (21,75)(22,68)(23,52)(24,34)(25,32)  };

   \legend{ $\alpha=14$, $\alpha=18$, $\alpha=22$, $\alpha=26$, $\alpha=30$, $\alpha =34$, $\alpha=38$, $\alpha=40$}
\end{axis}
\end{tikzpicture}
\qquad}

\fbox{\begin{tikzpicture}[scale=0.33]
\begin{axis}
	[
    title={(d). Identification rate for $n=700, m=96$},
    xlabel={\#Positives, $d$ },
    ylabel={Success rate\%},
    xmin=0, xmax=26,
    ymin=0, ymax=110,
    xtick={0,2,4,6,8,10, 12, 14, 16, 18, 20, 22, 24, 26},
    ytick={0,10,20,30, 40,50, 60,70, 80,90,100},
    legend pos=south west,
    ymajorgrids=true,
    grid style=dashed,
    ]
    
  \addplot  coordinates {
    (1,100)(2,100)(3,99)(4,95)(5,91)(6,91)(7,81)(8,75)(9,58)(10,48)(11, 37)(12, 28)(13,23)(14,16)(15,12)(16,3)(17,4)(18,5)(19,2)(20,0) (21,1)(22,0)(23,0)(24,0)(25,0)  };

 \addplot  coordinates {
    (1,100)(2,100)(3,100)(4,100)(5,98)(6,96)(7,86)(8,85)(9,75)(10,70)(11, 69)(12, 47)(13,48)(14,35)(15,32)(16,26)(17,18)(18,13)(19,10)(20,3) (21,3)(22,0)(23,0)(24,1)(25,0)  };

\addplot  coordinates {
    (1,100)(2,100)(3,100)(4,100)(5,100)(6,100)(7,100)(8,96)(9,98)(10,95)(11, 89)(12, 84)(13,75)(14,71)(15,68)(16,44)(17,48)(18,42)(19,35)(20,22) (21,22)(22,14)(23,3)(24,6)(25,2)  };

\addplot  coordinates {
    (1,100)(2,100)(3,100)(4,100)(5,100)(6,100)(7,100)(8,99)(9,100)(10,100)(11, 95)(12, 95)(13,83)(14,89)(15,76)(16,72)(17,76)(18,68)(19,42)(20,35) (21,41)(22,10)(23,11)(24,7)(25,11)  };

 \addplot [color = green, mark=*]  coordinates {
    (1,100)(2,100)(3,100)(4,100)(5,100)(6,100)(7,100)(8,100)(9,100)(10,100)(11, 98)(12, 96)(13,95)(14,93)(15,86)(16,79)(17,72)(18,65)(19,49)(20,51) (21,34)(22,35)(23,26)(24,12)(25,5)  };

 \addplot [color = orange, mark=*]  coordinates {
    (1,100)(2,100)(3,100)(4,100)(5,100)(6,100)(7,100)(8,100)(9,100)(10,99)(11, 98)(12, 98)(13,99)(14,99)(15,87)(16,80)(17,80)(18,72)(19,54)(20,48) (21,41)(22,22)(23,33)(24,21)(25,12)  };

 \addplot [color = cyan, mark=triangle]  coordinates {
    (1,100)(2,100)(3,100)(4,100)(5,100)(6,100)(7,100)(8,100)(9,100)(10,100)(11, 100)(12, 98)(13,100)(14,98)(15,95)(16,91)(17,88)(18,77)(19,70)(20,63) (21,47)(22,31)(23,24)(24,35)(25,16)  };

 \addplot [color = black, mark=triangle]  coordinates {
    (1,100)(2,100)(3,100)(4,100)(5,100)(6,100)(7,100)(8,100)(9,100)(10,100)(11, 100)(12, 100)(13,100)(14,98)(15,99)(16,92)(17,87)(18,87)(19,64)(20,63) (21,52)(22,35)(23,39)(24,21)(25,13)  };

   \legend{ $\alpha=16$, $\alpha=20$, $\alpha=24$, $\alpha=28$, $\alpha=32$, $\alpha =36$, $\alpha=40$, $\alpha=42$}
\end{axis}
\end{tikzpicture}
\qquad}
\fbox{\begin{tikzpicture}[scale = 0.33]
\begin{axis}
	[
    title={(e). Identification rate for $n=800, m=96$},
    xlabel={\#Positives, $d$ },
    ylabel={Success rate\%},
    xmin=0, xmax=26,
    ymin=0, ymax=110,
    xtick={0,2,4,6,8,10, 12, 14, 16, 18, 20, 22, 24, 26},
    ytick={0,10,20,30, 40,50, 60,70, 80,90,100},
    legend pos=south west,
    ymajorgrids=true,
    grid style=dashed,
    ]
    
  \addplot  coordinates {
    (1,100)(2,100)(3,100)(4,97)(5,91)(6,90)(7,77)(8,61)(9,63)(10,30)(11, 37)(12, 23)(13,17)(14,15)(15,8)(16,2)(17,1)(18,2)(19,1)(20,1) (21,0)(22,0)(23,0)(24,0)(25,0)  };

 \addplot  coordinates {
    (1,100)(2,100)(3,99)(4,97)(5,96)(6,94)(7,83)(8,79)(9,64)(10,47)(11, 52)(12, 39)(13,23)(14,22)(15,11)(16,10)(17,10)(18,1)(19,3)(20,0) (21,2)(22,1)(23,0)(24,0)(25,0)  };


\addplot  coordinates {
    (1,100)(2,100)(3,100)(4,100)(5,100)(6,99)(7,99)(8,100)(9,95)(10,88)(11, 89)(12, 74)(13,68)(14,69)(15,63)(16,36)(17,41)(18,31)(19,14)(20,13) (21,11)(22,5)(23,3)(24,3)(25,0)  };

\addplot  coordinates {
    (1,100)(2,100)(3,100)(4,100)(5,100)(6,100)(7,98)(8,98)(9,94)(10,91)(11, 85)(12, 84)(13,76)(14,76)(15,56)(16,57)(17,43)(18,38)(19,26)(20,17) (21,14)(22,14)(23,4)(24,3)(25,2)  };

 \addplot [color = green, mark=*]  coordinates {
    (1,100)(2,100)(3,100)(4,100)(5,100)(6,100)(7,100)(8,100)(9,99)(10,100)(11, 94)(12, 94)(13,86)(14,77)(15,66)(16,60)(17,61)(18,49)(19,36)(20,23) (21,21)(22,16)(23,8)(24,3)(25,7)  };

 \addplot [color = orange, mark=*]  coordinates {
    (1,100)(2,100)(3,100)(4,100)(5,100)(6,99)(7,100)(8,100)(9,99)(10,99)(11, 96)(12, 98)(13,88)(14,87)(15,70)(16,69)(17,68)(18,56)(19,51)(20,36) (21,22)(22,19)(23,10)(24,6)(25,2)  };

 \addplot [color = cyan, mark=triangle]  coordinates {
    (1,100)(2,100)(3,100)(4,100)(5,100)(6,100)(7,100)(8,100)(9,100)(10,100)(11, 99)(12, 99)(13,89)(14,97)(15,91)(16,77)(17,71)(18,64)(19,56)(20,39) (21,40)(22,16)(23,20)(24,11)(25,8)  };

 \addplot [color = black, mark=triangle]  coordinates {
    (1,100)(2,100)(3,100)(4,100)(5,100)(6,100)(7,100)(8,100)(9,100)(10,100)(11, 100)(12, 99)(13,95)(14,94)(15,93)(16,81)(17,83)(18,67)(19,53)(20,47) (21,39)(22,22)(23,13)(24,12)(25,12)  };

   \legend{ $\alpha=18$, $\alpha=22$, $\alpha=26$, $\alpha=30$, $\alpha=34$, $\alpha =38$, $\alpha=42$, $\alpha=46$}   
\end{axis}
\end{tikzpicture}
\qquad}
\fbox{\begin{tikzpicture}[scale=0.33]
\begin{axis}
	[
    title={(f). Identification rate for $n=1000, m=96$},
    xlabel={\#Positives, $d$ },
    ylabel={Success rate\%},
    xmin=0, xmax=26,
    ymin=0, ymax=110,
    xtick={0,2,4,6,8,10, 12, 14, 16, 18, 20, 22, 24, 26},
    ytick={0,10,20,30, 40,50, 60,70, 80,90,100},
    legend pos=south west,
    ymajorgrids=true,
    grid style=dashed,
    ]
    
  \addplot  coordinates {
    (1,100)(2,100)(3,100)(4,96)(5,91)(6,81)(7,76)(8,53)(9,54)(10,33)(11, 22)(12, 10)(13,13)(14,8)(15,3)(16,1)(17,1)(18,0)(19,0)(20,0) (21,0)(22,0)(23,0)(24,0)(25,0)  };

 \addplot  coordinates {
    (1,100)(2,100)(3,99)(4,98)(5,92)(6,79)(7,78)(8,70)(9,45)(10,42)(11, 26) (12, 22)(13,10)(14,9)(15,7)(16,2)(17,1)(18,1)(19,0)(20,0) (21,0)(22,0)(23,0)(24,0)(25,0)  };

\addplot  coordinates {
    (1,100)(2,100)(3,100)(4,100)(5,95)(6,95)(7,93)(8,86)(9,83)(10,74)(11, 60)(12, 44)(13,36)(14,22)(15,30)(16,18)(17,13)(18,3)(19,6)(20,1) (21,1)(22,0)(23,0)(24,0)(25,0)  };

\addplot  coordinates {
    (1,100)(2,100)(3,100)(4,100)(5,99)(6,97)(7,97)(8,93)(9,92)(10,79)(11, 71)(12, 71)(13,56)(14,47)(15,29)(16,26)(17,23)(18,17)(19,6)(20,6) (21,1)(22,3)(23,2)(24,0)(25,0)  };

 \addplot [color = green, mark=*]  coordinates {
    (1,100)(2,100)(3,100)(4,98)(5,98)(6,99)(7,98)(8,94)(9,90)(10,87)(11, 80)(12, 75)(13,48)(14,62)(15,43)(16,30)(17,23)(18,17)(19,10)(20,19) (21,5)(22,2)(23,1)(24,1)(25,0)  };

 \addplot [color = cyan, mark=triangle]  coordinates {
    (1,100)(2,100)(3,100)(4,100)(5,100)(6,100)(7,100)(8,99)(9,98)(10,92)(11, 90)(12, 84)(13,75)(14,69)(15,58)(16,41)(17,34)(18,35)(19,23)(20,16) (21,10)(22,3)(23,2)(24,1)(25,0)  };

 \addplot [color = orange, mark=*]  coordinates {
 (1,100)(2,100)(3,100)(4,100)(5,100)(6,99)(7,100)(8,97)(9,96)(10,94)(11, 95)(12, 83)(13,81)(14,79)(15,58)(16,51)(17,43)(18,37)(19,28)(20,18) (21,12)(22,7)(23,6)(24,0)(25,4)  };

 \addplot [color = black, mark=triangle]  coordinates {
     (1,100)(2,100)(3,100)(4,100)(5,100)(6,100)(7,100)(8,99)(9,99)(10,94)(11, 91)(12, 90)(13,80)(14,79)(15,59)(16,59)(17,51)(18,30)(19,25)(20,30) (21,12)(22,6)(23,8)(24,3)(25,1) };

   \legend{  $\alpha=22$, $\alpha=26$, $\alpha=30$, $\alpha=34$, $\alpha =38$, $\alpha=42$, $\alpha=46$, $\alpha=48$}
   
\end{axis}
\end{tikzpicture}
\qquad}
\caption{Success rate for identification of positives with $p$ up to $5\%$, $n=400, \cdots, 1000$, $m=96$ and $\alpha\leq 48$, in the noise free case.}
\label{m96}
\end{figure*}
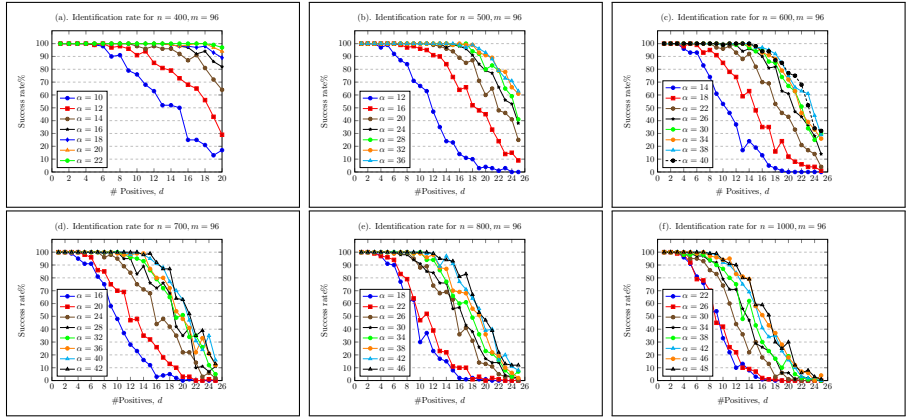

For $m=96$, the examined pool sizes that range from $10$ to $48$ when applicable. The positives ratio, denoted here by  $p$, ranges between $1\% -5\%$, and the tests were performed on sets of $n=400, 500, \ldots, 1000.$ In each setting, the minimum choice of $\alpha$ is derived from the applicability of the compressive sensing algorithm on the constructed matrix, while the maximum value of the pool size matches the maximum value tested for COVID-19 PCR pool designs~\cite{SNSWVS20}. Fig.~\ref{m96} describes the success rate  in restoring the defective items. The $x-$axis denotes the total number of the defectives $d$ in the set of $n$ elements. The success rate is examined while applying different values of $\alpha$. In this set of experiments, for each $n$ and each possible value of $d$ that is up to $5\%$ of the total number of elements $n$, the target vector $\hat{x}$ is chosen where the $d$ non-zero entries are chosen uniformly at random, while the symptom load of each non-zero entry is chosen uniformly over the real range $[0,1]$. For each choice of $n,d$ and $\alpha$, the success rate is evaluated over $100$ $m\times n$  different $\alpha-$RRD matrices and $100$ corresponding $\hat{x}$ for each design. For each $\hat{x}$ and design pair, we check if GTCS restored the vector $x$ correctly. We remind the reader that $x$ is a binary vector derived from $\hat{x}$ where its non-zero entries indicate the set of positive items. The success rate in Fig.~\ref{m96} reflects the number of trials in which exact restoration of $x$ was achieved over the total number of trials, i.e. $100$.

Our experiments results suggest that using the GTCS framework with $\alpha-$RRD design can reduce the number of tests dramatically. Morover, the results imply that there is a correlation between the value of $\alpha$ and the performance of the process; choosing higher values of $\alpha$ can increase the success rate in identifying the positives. This can be observed from the results summarized in Fig.~\ref{m96}. For example, for $n=400$ and  $p=5\%$, the success rate for $\alpha=10$ is bellow $20\%$ while it can be increased beyond $95\%$ when $\alpha=22$. This paradigm is reproduced in all test configurations  outlined in Fig.~\ref{m96}. 

In addition, for $n=400$ and positives rate $p=1\%$,  a success rate of $99\%$ is achieved for pool size as small as 10. This implies that we can test  up to $n=400$ samples with only $96$ tests when the positives rate is $1\%$ with success rate as high as $99\%$. When the positives rate rises to $5\%$, to achieve this success rate, $\alpha$ needs to be increased up to $28$. For $n=900$, designs with pool size $\alpha=48$ can be used to identify positives when $p=1\%$ using the same amount of tests. This indicates that the number of tests can be reduced by up to $10$ folds when  $\alpha =48$ for $n=900$.  For rates higher than that,  no value of $\alpha \leq 48$ could achieve the success rate of $99\%$.  Table~(\ref{m96bestrate}) summarizes the best (minimal) choice of $\alpha$ for $n$ and $p$ to achieve  success rate higher that $99\%$.

 \ignore{Table~(\ref{m96}) summarizes the optimal choice of $\alpha$ relative to $n$ and the positives rate  of the $\alpha-$RRD design when the number of the tests is fixed to $m=96$ for positives rates of $1\%\ldots 5\%$ and $n=400, \ldots, 900$. 
The data in Figure~(\ref{m96}) suggests that increasing the value of $\alpha$ increases the success rate. On the other hand, the data also suggests that the improvement that increasing $\alpha$ yields decays. }

\begin{table}[b!]
\centering
 \caption{Pool size per positives rate to achieve success rate $>99\%$ when $m=96.$}
 \begin{tabular}{| c| c |c| c| c |c |} 
 \hline
  \textbf{\#Items}&\multicolumn{5}{|c|}{\textbf{Positives rate $p$}} \\
 \cline{2-6}
&&&&&\\
  $n$& $\mathbf{1\%}$ & $\mathbf{2\%}$ & $\mathbf{3\%}$ & $\mathbf{4\%}$ & $\mathbf{5\%}$ \\ [0.5ex] 
 \hline\hline
  \textbf{400} & 10 & 14& 16 & 22& 28 \\ 
 \hline
  \textbf{500} & 16 & 20 & 24 &-&- \\
 \hline
 \textbf{600} & 20 & 26& -&-&-\\
 \hline
 \textbf{700} & 24 & 44  &-&-&-\\
  \hline
\textbf{800} & 38 &- &-&-& -\\ 
  \hline
\textbf{900} & 48 & -&-&-&-\\ [1ex] 
  \hline
 \end{tabular}

 \label{m96bestrate}
\end{table}

For $m=384$, we tested the performance of GTCS with RRD designs when $\alpha$ ranges from $8$ up to $48$ and $n$ ranges from $800$ up to $2000$. In this set, for each choice of $n$ and $\alpha$, the success rate is evaluated on $10$ random designs. Table~(\ref{m384bestrate}) lists the minimal choice of $\alpha$ that is required to achieve high success rate ($\geq 99\%$) for each value of $n$ and positives ratio $p$. 

 In addition, we tested a configuration with $n=4000$, $\alpha=48$ and positives rate up to $p=1.5\%$. For this configuration, we  achieved a success rate of $100\%$ for our choice of $\alpha = 48$ and positives rate $p=1.3\%$; this is a 10 fold improvement in the number of tests required to test a group of magnitude that reaches $4000$ samples.
 
 Both sets of experiments were performed on a simulator specifically programmed to implement the algorithms used in these tests. The simulator can be easily configured to provide  an $\alpha-$RRD design for any choice of $\alpha, n,m$ and $p$. When these parameters are provided, the simulator generates a configurable number of test designs among which it chooses the one that performs the best on its auto-generated synthetic datasets.    

\begin{table*}[t!]
\centering
\caption{Minimal pool size ($\alpha$) per positives rate to achieve success rate $>99\%$ when $m=384$}
 \begin{tabular}{|c | c| c |c| c| c |c|c |c |c|c|} 
 \hline
 \textbf{\#Items}&\multicolumn{10}{|c|}{\textbf{Positives rate $p$}} \\
 \cline{2-11} 

 &  & & & & &&& & & \\ 
\textbf{\textit{n}}& $\mathbf{1\%}$ & $\mathbf{2\%}$ & $\mathbf{3\%}$ & $\mathbf{4\%}$ & $\mathbf{5\%}$  & $\mathbf{6\%}$ & $\mathbf{7\%}$ & $\mathbf{8\%}$ & $\mathbf{9\%}$& $\mathbf{10\%}$\\ [0.5ex] 
 \hline\hline
 \textbf{800} & 8 &8 &8 &8 &8 &8 &8 &8 &8 &8 \\ 
 \hline
 \textbf{900} & 10 & 10 & 10 & 10 & 10 & 10 & 10 & 10 & 10 & 10  \\  
 \hline
\textbf{1000} & 12& 12& 12& 12& 12& 12& 12& 12& 12& 12\\
 \hline
 \textbf{1100} & 14& 14& 14& 14& 14& 14& 14& 14& 14& 14\\
  \hline
 \textbf{1200} & 16& 16& 16& 16& 16& 16& 16& 16 & 18 & 22 \\ 
  \hline
 \textbf{1300} &18&18&18&18&18&18&18&20&22&-\\ 
 \hline
 \textbf{1400} & 20& 20& 20& 20& 20& 20&20 & 26 & - & - \\ 
  \hline
\textbf{1500} & 22& 22& 22& 22& 22& 22& 28& - &- & - \\ 
  \hline
 \textbf{1600 }& 24& 24& 24& 24& 24& 24& -& - & - & - \\ 
  \hline
  \textbf{1700 }& 26& 26& 26& 26& 26& 28& -& - & - & - \\ 
  \hline
  \textbf{1800 }& 28& 28& 28& 28& 28& -& -& - & - & - \\ 
  \hline
  \textbf{1900} & 30& 30& 30&30& -& -& -& - & - & - \\ 
  \hline
  \textbf{2000} & 32& 32& 32& 32& -& -& -& - & - & -\\ 
  \hline
  \textbf{4000} & 48& -& -& -& -& -& -& - & - & -\\ [1ex] 
  \hline
 \end{tabular}
  \label{m384bestrate}
\end{table*}

\section{Summary and Conclusions }

In this paper, we suggested a new random pooling design $\alpha-$RRD. This design can be used as part of the GTCS paradigm in order to build a single-round non-adaptive group testing protocol to exactly identify positives within a large set of elements.  The complexity of generating $\alpha-$RRD design is polynomial in the dimensions of the design and can be easily generated for any dimensions $m$ and $n$. Given the parameters $m, n$ and positives rate, the best choice for the parameter $\alpha$ can be concluded using computer simulations. Moreover, since random sensing matrices can perform well with compressive-sensing algorithms, the GTCS paradigm can be further tested with other well-known group testing random designs such as RID, RrSD, RsSD and Transversal design~\cite{BHHZ19}. Similarly, other compressive sensing algorithms can be applied too.  We performed the tests in this paper on noiseless synthetic data sets, however, from practical point of view, it is encouraged to check the performance on a noisy data or even real-life data while applying standard methods designed for coping with inaccuracies and errors dictated by testing processes.


\bibliographystyle{plain}
\bibliography{Ref}
\vspace{12pt}

\end{document}